\documentclass[conference]{IEEEtran}
\usepackage[left=0.67in,right=0.67in,top=0.64in,bottom=0.70in]{geometry}
\usepackage[T1]{fontenc}
\usepackage[latin9]{inputenc}
\usepackage{amsthm}
\usepackage{amsmath} 
\usepackage{graphicx} 
\usepackage{color} 
\usepackage{cite}
\usepackage{algpseudocode}
\usepackage{algorithm}
\usepackage{amssymb}
\usepackage{subfigure}
\usepackage{esint}
\usepackage{algpseudocode}
\usepackage{epstopdf}
\usepackage{multirow}
\usepackage{makecell}
\usepackage{float}
\usepackage{lipsum}
\usepackage{balance}
\usepackage{hhline}
\usepackage{arydshln}
\usepackage{epstopdf}

\newtheorem{remark}{Remark}

\theoremstyle{plain}
\theoremstyle{plain}
\newtheorem{theorem}{Theorem}

\newcommand{\comment}[1]{}

\IEEEoverridecommandlockouts

\allowdisplaybreaks

\begin{document}

\title{Unequal Error Protection Achieves Threshold Gains on BEC and BSC via Higher Fidelity Messages\vspace{-0.2em}}

\author{
    \IEEEauthorblockN{Beyza Dabak\IEEEauthorrefmark{1}, Ahmed Hareedy\IEEEauthorrefmark{1}, Alexei Ashikhmin\IEEEauthorrefmark{2}, and Robert Calderbank\IEEEauthorrefmark{1}}
    \IEEEauthorblockA{\IEEEauthorrefmark{1}Electrical and Computer Engineering Department, Duke University, Durham, NC 27708 USA \\ \IEEEauthorrefmark{2}Mathematics and Algorithms of
    Communications Department, Nokia Bell Labs, Murray Hill, NJ 07974 USA \\ beyza.dabak@duke.edu, ahmed.hareedy@duke.edu, alexei.ashikhmin@nokia-bell-labs.com, and robert.calderbank@duke.edu\vspace{-0.5em}
   }
}
\maketitle
\begin{abstract}

Because of their capacity-approaching performance, graph-based codes have a wide range of applications, including communications and storage. In these codes, unequal error protection (UEP) can offer performance gains with limited rate loss. Recent empirical results in magnetic recording (MR) systems show that extra protection for the parity bits of a low-density parity-check (LDPC) code via constrained coding results in significant density gains. In particular, when UEP is applied via more reliable parity bits, higher fidelity messages of parity bits are spread to all bits by message passing algorithm, enabling performance gains. Threshold analysis is a tool to measure the effectiveness of a graph-based code or coding scheme. In this paper, we provide a theoretical analysis of this UEP idea using extrinsic information transfer (EXIT) charts in the binary erasure channel (BEC) and the binary symmetric channel (BSC). We use EXIT functions to investigate the effect of change in mutual information of parity bits on the overall coding scheme. We propose a setup in which parity bits of a repeat-accumulate (RA) LDPC code have lower erasure or crossover probabilities than input information bits. We derive the a-priori and extrinsic mutual information functions for check nodes and variable nodes of the code. After applying our UEP setup to the information functions, we formulate a linear programming problem to find the optimal degree distribution that maximizes the code rate under the decoding convergence constraint. Results show that UEP via higher fidelity parity bits achieves up to about $17\%$ and $28\%$ threshold gains on BEC and BSC, respectively.

\end{abstract}

\section{Introduction}\label{sec_intro}

Low-density parity-check (LDPC) codes, which are collectively called graph-based codes, are a family of error-correcting codes (ECC) that were introduced by Gallager \cite{gallager} in 1962. Three decades later, with the advances of circuit design and their decoding algorithms, LDPC codes were revisited and new code constructions, including repeat-accumulate (RA) \cite{divsalar} and irregular repeat-accumulate (IRA) codes \cite{jin_ira}, were proposed. Today, graph-based codes have applications in many areas including wireless communication and data storage.

In ECC, unequal error protection (UEP) is used in applications where some of the channel bits are more sensitive to error or where error in a specific feature of the data is more costly than others, i.e., data has unequal value to the user \cite{calderbank_uep, katsman_uep}. Codes with unequal protection of bits, and higher protection of input information bits over parity bits offer performance gains \cite{wolf_uep,furzun_uep}. In this paper, we apply UEP with higher protection of the parity bits over the input information bits of an RA code to limit rate loss, and we demonstrate threshold gains.\footnote{Although in our UEP setup we protect some \textit{data} bits without any additional error-correction, we stick to the more common nomenclature of unequal \textit{error} protection. Moreover, error here refers to flips/erasures.}

The idea of applying higher protection on parity bits was introduced in \cite{ahh_loco} in the context of data storage. The authors introduced lexicographically-ordered constrained (LOCO) codes, which significantly mitigate inter-symbol interference (ISI) in magnetic recording (MR) systems. In their model, parity bits of a spatially-coupled (SC) LDPC code are encoded via a LOCO code, and up to $20\%$ density gain, with limited rate loss, is achieved compared with using the LDPC code only. Protecting parity bits solely via a LOCO code achieves a significant rate-density gain trade-off. The same UEP idea can also be used to limit speed loss in Flash and other applications where the constrained code rate is already high \cite{ahh_qaloco, ahh_general}. This UEP setup is successful because when parity bits have higher fidelity messages, e.g., log-likelihood ratios (LLRs), those reliable messages are diffused into all bits during message passing \cite{ahh_loco}. Thus, the decoder effectively experiences a higher signal-to-noise ratio (SNR) compared to the uniform case. The empirical results presented in \cite{ahh_loco} are the motivation behind this paper, which is to demonstrate the threshold gains of this UEP setup theoretically. We start with the binary erasure channel (BEC) and the binary symmetric channel (BSC), and subject parity bits to lower erasure and crossover probabilities, respectively, compared to input information bits. With this setup, we ensure obtaining the closest model to achieving high reliability on parity bits which was done by constrained coding in \cite{ahh_loco}.

We use extrinsic information transfer (EXIT) charts \cite{brink_cid} for the threshold analysis of RA LDPC codes with our idea of UEP via more reliable parity bits on BEC and BSC. EXIT charts are a tool to visualize the asymptotic performance and predict convergence behavior of iterative decoders \cite{brink_cid, alexei_bec}. EXIT charts plot \textit{average extrinsic information} coming out of the decoder as a function of \textit{average a-priori information} going into the decoder during the iterations. In the literature, EXIT charts were used in the design of RA codes that are capacity-approaching \cite{brink_ra}. EXIT functions were derived for BEC, and models for the decoding of RA and general LDPC codes were introduced in \cite{alexei_bec}. In \cite{sharon_bms}, methods to obtain EXIT functions for binary-input memoryless symmetric channels were developed through an alternative pseudo-MAP decoder.

In this paper, we formulate and solve a linear programming (LP) problem to find the optimal degree distribution of the RA code, which maximizes the rate given the erasure or crossover probabilities such that decoding convergence is guaranteed on the EXIT chart, for both the UEP and the uniform setups. For our UEP setup with more reliable parity bits, EXIT functions are used to investigate how the change in the mutual information of parity bits affects the behavior of the overall coding scheme. We discuss an alternative derivation of the EXIT functions for BEC from a combined channel perspective, and using the decoding model introduced in \cite{alexei_bec}, we derive EXIT functions of variable nodes (VNs) and check nodes (CNs) of an RA LDPC code for BSC. Our experimental results demonstrate the effectiveness of this UEP idea as it achieves up to about $17\%$ and $28\%$ threshold gains on BEC and BSC, respectively. The ideas and results we present in this paper are the first step towards developing the UEP theoretical framework for modern data storage systems.

The rest of the paper is organized as follows. In Section~\ref{sec_prelim}, we discuss the preliminaries. In Section~\ref{sec_steps}, we introduce our theoretical methodology and derive the (LP) problem for our UEP idea. In Section~\ref{sec_bec}, we apply this methodology to BEC, and show threshold gains. In Section~\ref{sec_bsc}, we do the same for BSC. Section~\ref{sec_conc} concludes the paper.

\section{Preliminaries}\label{sec_prelim}

\begin{figure}
\vspace{-0.3em}
\centering
\includegraphics[trim={2.1in 3.5in 2.2in 1.7in},clip,width=3.7in]{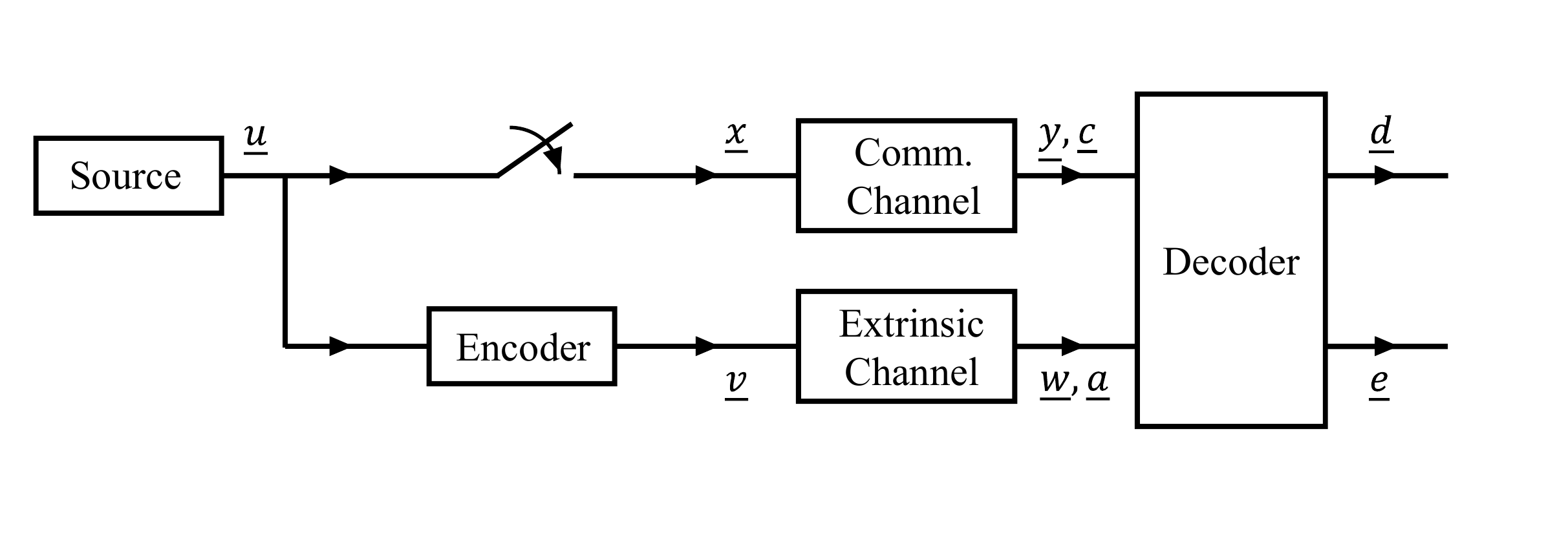}
\vspace{-2.7em}
\caption{Decoding model for EXIT analysis}
\label{fig_1}
\vspace{-0.8em}
\end{figure}

We use the decoding model shown in Fig.~\ref{fig_1} for EXIT chart analysis. A binary-symmetric source produces bits that take $0$ or $1$ value with equal probability. There exist $P(\underline{y}|\underline{x})$ \textit{communication channel} and $P(\underline{w}|\underline{v})$ \textit{extrinsic channel} with output vectors $\underline{y}$ and $\underline{w}$ (noisy versions of the inputs $\underline{x}$ and $\underline{v}$) and LLRs $\underline{c}$ and $\underline{a}$, respectively. Fig.~\ref{fig_1} models iterative decoding where the extrinsic channel, which is actually an artificial channel, models extrinsic information coming from the previous decoding iteration \cite{alexei_bec}. The decoder uses outputs of both channels $\underline{y}$ and $\underline{w}$ to calculate \textit{a-posteriori} and \textit{extrinsic} LLRs $\underline{d}$ and $\underline{e}$ of $\underline{v}$, respectively. See \cite{alexei_bec} for more details.

We investigate the threshold gains of higher protection of the~parity bits of an RA code using EXIT chart analysis. When considering VNs of the RA code, the switch in Fig.~\ref{fig_1} is closed, resulting in $\underline{u} = \underline{x}$, $\underline{u}$ is one bit, and the Encoder is a repetition code with length $d_{\textup{v}}$ (VN degree). Whereas when considering the CNs of the RA code, the switch in Fig.~\ref{fig_1} is open, and the Encoder is a single parity-check (SPC) code with length $d_{\textup{c}}$ (CN degree) \cite{alexei_bec}. We investigate the effect of UEP on RA codes in two setups. In the first setup, both communication and extrinsic channels are BECs with erasure probabilities $q$ and $p$, respectively. In the second, both channels are BSCs with crossover probabilities $\epsilon$ and $\delta$, respectively.

In the model of iterative decoding, extrinsic LLR $e_{i}$ at the decoder output in one iteration re-enters the decoder as a-priori LLR $a_{i}$ after passing through an interleaver in the next iteration \cite{alexei_bec, brink_cid}. This is consistent with the  modelling of the extrinsic channel discussed above. Next, EXIT functions are defined. Let $m$ be the length of $\underline{v}$, $\underline{w}$, $\underline{a}$, and $\underline{e}$. The \textit{average a-priori information} $I_{\textup{A}}$ going into the decoder is then \cite{alexei_bec}:
\begin{equation} \label{IA}
I_{\textup{A}} = \frac{1}{m}\sum_{i=1}^{m} I(V_{i};A_{i}) = I(V_{1};A_{1}).
\end{equation} 
The second equality follows from the observation that $V_{i}$, for all $i$, have the same distribution and that the extrinsic channel is memoryless and time invariant. The \textit{average extrinsic information} $I_{\textup{E}}$ coming out of the decoder is \cite{alexei_bec}:
\begin{equation} \label{IE}
I_{\textup{E}} = \frac{1}{m}\sum_{i=1}^{m} I(V_{i};E_{i}) = I(V_{1};E_{1}) = I(V_{1};\underline{Y},\underline{A}_{[1]}),
\end{equation} 
where an a-posteriori probability (APP) decoder is assumed. We write random variables with upper case letters, their realizations with lower case letters. $\underline{A}_{[i]}$ denotes vector $\underline{A}$ with the $i$th entry removed. The third equality in (\ref{IE}) follows from the proposition proved in \cite{alexei_bec} for APP decoders with extrinsic message passing. This allows the extrinsic information to be defined as mutual information between input of the extrinsic channel and the inputs of the decoder instead of the extrinsic LLR at the output of the decoder. An EXIT chart plots extrinsic information as a function of a-priori information. \cite{brink_cid}.

\section{Methodology for UEP Analysis}\label{sec_steps}

For our UEP setup, we use EXIT functions to see how a change in the mutual information of parity bits affects the behavior of the overall coding scheme. In this section, we illustrate the code construction and our methodology with steps which are applied to BEC and BSC in Sections~\ref{sec_bec} and~\ref{sec_bsc}.

To guarantee the diffusion of higher fidelity messages from parity to input information bits, a specific property is required in the code construction. That is, for each VN representing an input information bit, there exists at least one VN representing a parity bit connected to the first through a CN. This property is satisfied in RA LDPC codes with parity-check matrix $H$ of the form $H = [J\textup{ }P]$, where $P$ is $(n-k)\times(k+1)$ sparse matrix and $J$ is $(n-k)\times(n-k-1)$ matrix of the form:
\begin{equation}
J = 
\begin{bmatrix}
1 & 0 &   &   & &...& 0 \\
1 & 1 & 0 &   & &...& 0 \\
0 & 1 & 1 & 0  & &...& 0 \\
0 & 0 & 1 & 1 &0 &...& 0 \\
  &   &   & ...  & &&   \\
0 & ...  &  &   & 0 & 1 & 1 \\
0 & ...  &  &   &  & 0 & 1 
\end{bmatrix}.
\end{equation}
Assuming that the first column of $P$ has weight $2$, and it is linearly independent from the columns of J, the first $(n-k)$ bits of the RA codeword (corresponding to the columns of $J$ and the first column of $P$) can be considered parity bits, whereas the last $k$ bits can be considered input information bits. Also assuming $d_{\textup{c}} > 3$, each parity bit (except the first, $(n-k-1)$-th, and the last bits) is connected to another parity bit and at least $d_{\textup{c}}-3$ input information bits, which satisfies the required property. ($P$ has no $\underline{0}$ columns.)

With this RA code construction, we can now derive EXIT functions for our UEP setup via the following steps. Let the communication and extrinsic channels have error probabilities $\sigma$ and $\beta$, respectively.

\noindent\textbf{Step~1:} Consider the VNs of the RA code. First, derive the extrinsic information $I_{\textup{E,v}}$ as a function of the a-priori information $I_{\textup{A,v}}$ for the VNs without considering UEP, i.e., assuming both input information and parity bits are transmitted via a communication channel with fixed error probability $\sigma$. This step assumes fixed $d_{\textup{v}}$ for simplicity.

\noindent\textbf{Step~2:} Consider the CNs of the RA code. Derive the extrinsic information $I_{\textup{E,c}}$ as a function of the a-priori information $I_{\textup{A,c}}$ without considering UEP. Next, derive the inverse EXIT function of CNs.\footnote{Always $I_{\textup{A,v}} =  I_{\textup{A,c}}$. Thus, we use  $I_{\textup{A}}$ notation in the rest of the paper.} We adopt fixed $d_{\textup{c}}$ in our RA code construction.

\noindent\textbf{Step~3:} We are now ready to apply unequal protection on parity and input information bits. Let all CNs have degree $d_{\textup{c}}$. Let $\lambda_i$ be the fraction of branches (edges) connected to VNs of degree $i$, which we refer to as the {\textit{degree distribution}}\cite{jin_ira}.\footnote{In this paper, we refer to an edge that is adjacent to a node as "connected" to the node. Here, we mean they are directly connected.} Let $N$ be the total number of branches. Let $\lambda_2 = a + b$, where $a$ is the fraction of branches connected to $(n-k)$ VNs corresponding to $(n-k)$ parity bits. It follows from the RA code construction discussed earlier that $n-k = N\cdot \frac{1}{d_{\textup{c}}} = N\cdot \frac{a}{2} \implies a = \frac{2}{d_{\textup{c}}}$, where $n-k$ is the number of CNs. Thus we have,
\begin{equation} \label{constraint1}
    \frac{2}{d_{\textup{c}}} + b + \sum_{i \geq 3}\lambda_i = 1,
\end{equation}
which is the first constraint of the linear programming (LP) problem to be explained in the upcoming step.

Let $\sigma_{1}$ and $\sigma_{2}$ be the error probabilities of parity bits and input information bits transmitted through the communication channel, respectively. We now derive $I^{*}_{\textup{E,v}}$ for the  UEP setup:
\begin{align} \label{general_Iev}
    I^{*}_{\textup{E,v}} &= a \cdot I_{\textup{E,v}}(\sigma = \sigma_1, d_{\textup{v}} = 2) + b \cdot I_{\textup{E,v}}(\sigma = \sigma_2, d_{\textup{v}} = 2) \nonumber \\ &\hspace{+1.0em}+  \sum_{i \geq 3}\lambda_i \cdot I_{\textup{E,v}}(\sigma = \sigma_2, d_{\textup{v}} = i),
\end{align}
which is a weighted sum over branches. Note that we extend the arguments of $I_{\textup{E,v}}$ in (\ref{general_Iev}) for clarity.\footnote{The remaining information equations ($I_{\textup{E,c}}$ and $I_{\textup{A}}$) are same for unequal and uniform error protection (also same notation). A-priori information for both VNs and CNs depend only on the input and output of extrinsic channel, not the communication channel. Same for extrinsic information when considering CNs due to the open switch in Fig.~\ref{fig_1}.} Note also that (\ref{general_Iev}) is used for the uniform protection setup as well by setting $\sigma_{1} = \sigma_{2}$.

\noindent\textbf{Step~4:} In this step, we formulate an LP problem. Iterative decoding will be successful, i.e., convergence occurs, if the EXIT function of VNs lies above and does not intersect with the inverse of EXIT function of CNs \cite{alexei_bec}, i.e.,
\begin{equation} \label{constraint2}
    I^{*}_{\textup{E,v}}(I_{\textup{A}}) > I^{-1}_{\textup{E,c}}(I_{\textup{A}}),  \textup{ } I_{\textup{A}} \in (0,1), 
\end{equation}
which is the second constraint of the LP problem.

We calculate the code rate $R$ as follows:
\begin{align} \label{rate}
    R = 1- \frac{\frac{1}{d_{\textup{c}}}}{\frac{a}{2} + \frac{b}{2} + \sum_{i \geq 3} \frac{\lambda_i}{i}} = 1- \frac{\frac{1}{d_{\textup{c}}}}{\frac{1}{d_{\textup{c}}} + \frac{b}{2} + \sum_{i \geq 3} \frac{\lambda_i}{i}}
\end{align}

We then formulate the following LP problem for finding optimal degree distribution that maximizes the rate of the code under code construction and EXIT convergence constraints, (\ref{constraint1}) and (\ref{constraint2}), for pre-determined $\sigma_{1}$, $\sigma_{2}$ error probabilities:
\begin{align} \label{LP}
    \nonumber \textbf{maximize } &\; \frac{b}{2} + \sum_{i \geq 3} \frac{\lambda_i}{i} \\ \nonumber
    \textbf{subject to } &\; \frac{2}{d_{\textup{c}}} + b + \sum_{i \geq 3}\lambda_i = 1, \nonumber \\
    & I^{*}_{\textup{E,v}}(I_{\textup{A}}) > I^{-1}_{\textup{E,c}}(I_{\textup{A}}),  \textup{ } I_{\textup{A}} \in (0,1). 
\end{align}

This LP problem is derived in order to investigate the gains of unequal error protection. We solve the LP problem numerically using a software program. Given the channel probabilities, the solution of this optimization problem gives the degree distribution that achieves the highest rate. How to use the LP solution to obtain the threshold gains is discussed in the following sections.

\section{Unequal Error Protection on BEC}\label{sec_bec}

Let the communication and extrinsic channels in Fig.~\ref{fig_1} be BECs with erasure probabilities $q$ and $p$, respectively. In this section, we apply the steps outlined in Section~\ref{sec_steps} and discuss the results of applying unequal error protection on parity and input information bits of the RA code for BEC.

\noindent\textbf{\underline{Step~1:}} When considering the VNs of the RA code, $\underline{u} = \underline{x}$, and the Encoder is a repetition code with length $d_{\textup{v}}$. From \cite{alexei_bec} (see also the intuitive explanation after Step 2), for fixed $d_{\textup{v}}$,
\begin{align}
    & I_{\textup{A}} =  I(V_{1};A_{1}) = 1-p, \\ 
    & I_{\textup{E,v}} = 1-qp^{d_{\textup{v}}-1}, \label{bec_Iev} \\
    & I_{\textup{E,v}}(I_{\textup{A}}) = 1-q(1-I_{\textup{A}})^{d_{\textup{v}}-1}. \label{fcn_bec_Iev}
\end{align}
 
\noindent\textbf{\underline{Step~2:}} When considering the CNs of the RA code, the switch on the top branch is open, and the Encoder is an SPC code with length $d_{\textup{c}}$. From \cite{alexei_bec} (see also the intuitive explanation after Step 2) and for fixed $d_{\textup{c}}$,
\begin{align}
    & I_{\textup{A}} =  I(V_{1};A_{1}) = 1-p, \\
    & I_{\textup{E,c}} = (1-p)^{d_{\textup{c}}-1}, \label{bec_Iec} \\
    & I_{\textup{E,c}}(I_{\textup{A}}) = (I_{\textup{A}})^{d_{\textup{c}}-1}, \label{fcn_bec_Iec} \\
    & I^{-1}_{\textup{E,c}}(I_{\textup{A}}) = (I_{\textup{A}})^{\frac{1}{d_{\textup{c}}-1}}. \label{fcn_bec_Iec_inv}
\end{align}

To intuitively explain (\ref{bec_Iec}), we can think of a combined BEC setup. Let us consider the CNs side first. In order that a CN sends a correct message to a VN, the messages from all other $d_{\textup{c}}-1$ VNs must be correct. The probability of getting correct information from a VN straight from a BEC($p$) \textit{channel} is $1-p$. Note that here, a \textit{channel} refers to the extrinsic channel, not the communication channel. The probability that all those $d_{\textup{c}}-1$ VNs send correct messages is $p_{\textup{correct}} = (1-p)^{d_{\textup{c}}-1}$. Thus, the probability of receiving wrong (erased) information is $p_{\textup{wrong}} = 1-p_{\textup{correct}} = 1-(1-p)^{d_{\textup{c}}-1}$, which is actually the erasure probability ($p_{\textup{erasure}}$) of the new combined channel. Hence, the mutual information under equiprobable inputs (the capacity) of the combined BEC is $I_{\textup{E,c}} = 1-p_{\textup{erasure}} = 1-(1-(1-p)^{d_{\textup{c}}-1}) = (1-p)^{d_{\textup{c}}-1}$, which is (\ref{bec_Iec}).

The same logic can be applied to the VNs side to get (\ref{bec_Iev}), where the combined channel erasure probability will be $qp^{(d_{\textup{v}}-1)}$ this time. In order that a VN sends a wrong message to a CN, the information at that VN has to be erased by the communication channel, which is the first event, and all messages coming to that VN from the other $d_{\textup{v}}-1$ CNs are also erased, which is the second event. The probability of the first event is $q$, and the probability of the second event is $p^{d_{\textup{v}}-1}$. Note that any CN with non-erased information suffices to fix the erasure at that VN (different from BSC, see Section~\ref{sec_bsc}). Hence, $qp^{(d_{\textup{v}}-1)}$ is the erasure probability $p_{\textup{erasure}}$ of the new combined channel. Thus, $I_{\textup{E,v}} = 1-p_{\textup{erasure}} = 1-qp^{(d_{\textup{v}}-1)}$, which is (\ref{bec_Iev}). 

\begin{table*}
\caption{Threshold Gains of UEP at Various Erasure Probabilities When LP is Solved for Uniform and Unequal Error Protection on BEC}
\vspace{-0.5em}
\centering

\begin{tabular}{|c|c|c|c|c|c!{\vrule width 0.85pt}c|c|c|c|c|c|}
\hline
\multicolumn{6}{|c!{\vrule width 0.85pt}}{\makecell{\textbf{First method:} LP solved for uniform protection}} & \multicolumn{6}{c|}{\makecell{\textbf{Second method:} LP solved for unequal protection}} \\
\hline
Rate & $q_{\textup{uniform}}$ & $q'_1$ & {$q'_2$} & $q'_{\textup{avg}}$ & \textbf{Gain} & Rate & $q_1$ & {$q_2$} & $q_{\textup{avg}}$ & $q'_{\textup{uniform}}$  & \textbf{Gain} \\
\hline
$0.6316$ & $0.28$ & $0.003$ & $0.488$ & $0.3093$ & $10.5\%$ & $0.6376$ & $0.050$ & $0.500$ & $0.3369$ & $0.2883$ & $16.9\%$ \\
\hline
$0.6430$ & $0.25$ & $0.002$ & $0.426$ & $0.2746$ & $9.9\%$ & $0.6644$ & $0.080$ & $0.430$ & $0.3126$ & $0.2739$ & $14.1\%$ \\
\hline
$0.7237$ & $0.20$ & $0.003$ & $0.299$ & $0.2172$ & $8.6\%$ & $0.7347$ & $0.080$ & $0.300$ & $0.2416$ & $0.2180$ & $10.8\%$ \\
\hline
$0.7853$ & $0.14$ & $0.003$ & $0.188$ & $0.1483$ & $5.9\%$ & $0.7876$ & $0.090$ & $0.210$ & $0.1845$ & $0.1750$ & $5.4\%$ \\
\hline
$0.8527$ & $0.10$ & $0.004$ & $0.122$ & $0.1046$ & $4.6\%$ & $0.8526$ & $0.004$ & $0.122$ & $0.1046$ & $0.1009$ & $3.7\%$ \\
\hline
\end{tabular}

\label{table1}
\vspace{-0.25em}
\end{table*}

\noindent\textbf{\underline{Step~3:}} We now apply the UEP setup explained in the previous section to derive $I^{*}_{\textup{E,v}}$. Parity and input information bits transmitted through the communication channel face erasure probabilities $q_{1}$ and $q_{2}$, respectively. Using (\ref{general_Iev}) and (\ref{fcn_bec_Iev}), the EXIT function is derived to be:
\begin{align} \label{UEP_bec_fcn}
    I^{*}_{\textup{E,v}}(I_{\textup{A}}) &= a \cdot (1-q_{1}(1-I_{\textup{A}})) + b \cdot (1-q_{2}(1-I_{\textup{A}})) \nonumber \\ &\hspace{+1.0em}+  \sum_{i \geq 3}\lambda_i \cdot (1-q_{2}(1-I_{\textup{A}})^{i-1}).
\end{align}
\vspace{-0.6em}

\noindent\textbf{\underline{Step~4:}} Substituting (\ref{fcn_bec_Iec_inv}) and (\ref{UEP_bec_fcn}) in (\ref{constraint2}) completes the LP problem in (\ref{LP}). We now solve this LP problem to find the optimal degree distribution. That is, we maximize the rate of the code in (\ref{rate}), subject to degree distribution constraint (\ref{constraint1}) and EXIT convergence constraint (\ref{constraint2}).
\newline

We can find the solution ($a$, $b$ and $\lambda$'s) of the LP problem numerically using a software program. We implement two methods to investigate the threshold gains of our UEP idea over uniform protection in our software. In the first method, we mimic the approach adopted in data storage devices. In particular, the graph-based code is designed and optimized assuming all codeword bits will have the same protection \cite{ahh_loco,ahh_md}. As the device ages, the constrained code can be applied to the parity bits only, resulting in UEP and achieving density/lifetime gains \cite{ahh_loco}. Thus, in our first setup, we solve the LP problem for the uniform protection case, i.e, $q_1 = q_2 = q_{\textup{uniform}}$, and find the optimal degree distribution. An appropriate $d_{\textup{c}}$ is chosen according to $q_{\textup{uniform}}$. A code with this degree distribution has \textit{threshold} $q_{\textup{uniform}}$. After solving the LP problem for the uniform setup, we apply higher protection on parity bits (with $q'_1<q'_2$) for the same code construction (same $d_{\textup{c}}$ and degree distribution, thus same rate). We search for the ($q'_1, q'_2$) combination with the highest average that satisfies the EXIT convergence constraint in (\ref{constraint2}) using (\ref{fcn_bec_Iec_inv}), and (\ref{UEP_bec_fcn}) obtained in Step 3. We calculate the average erasure probability ($q'_{\textup{avg}}$) and the percentage threshold gains ($TG$) as follows: 
\begin{align}\label{eqn_qavg}
    &q'_{\textup{avg}} = q'_1(1-R) + q'_2(R), \\ \label{eqn_tg}
    &TG = \frac{q'_{\textup{avg}} - q_{\textup{uniform}}}{ q_{\textup{uniform}}} \times 100\%. 
\end{align}
Numerical results demonstrating the UEP threshold gains are illustrated in the left panel of Table~\ref{table1}.

\begin{figure}
\vspace{-1.0em}
\centering
\hspace{+0.5em}\includegraphics[trim={1.3in 0.5in 1.3in 0.5in},width=3.4in]{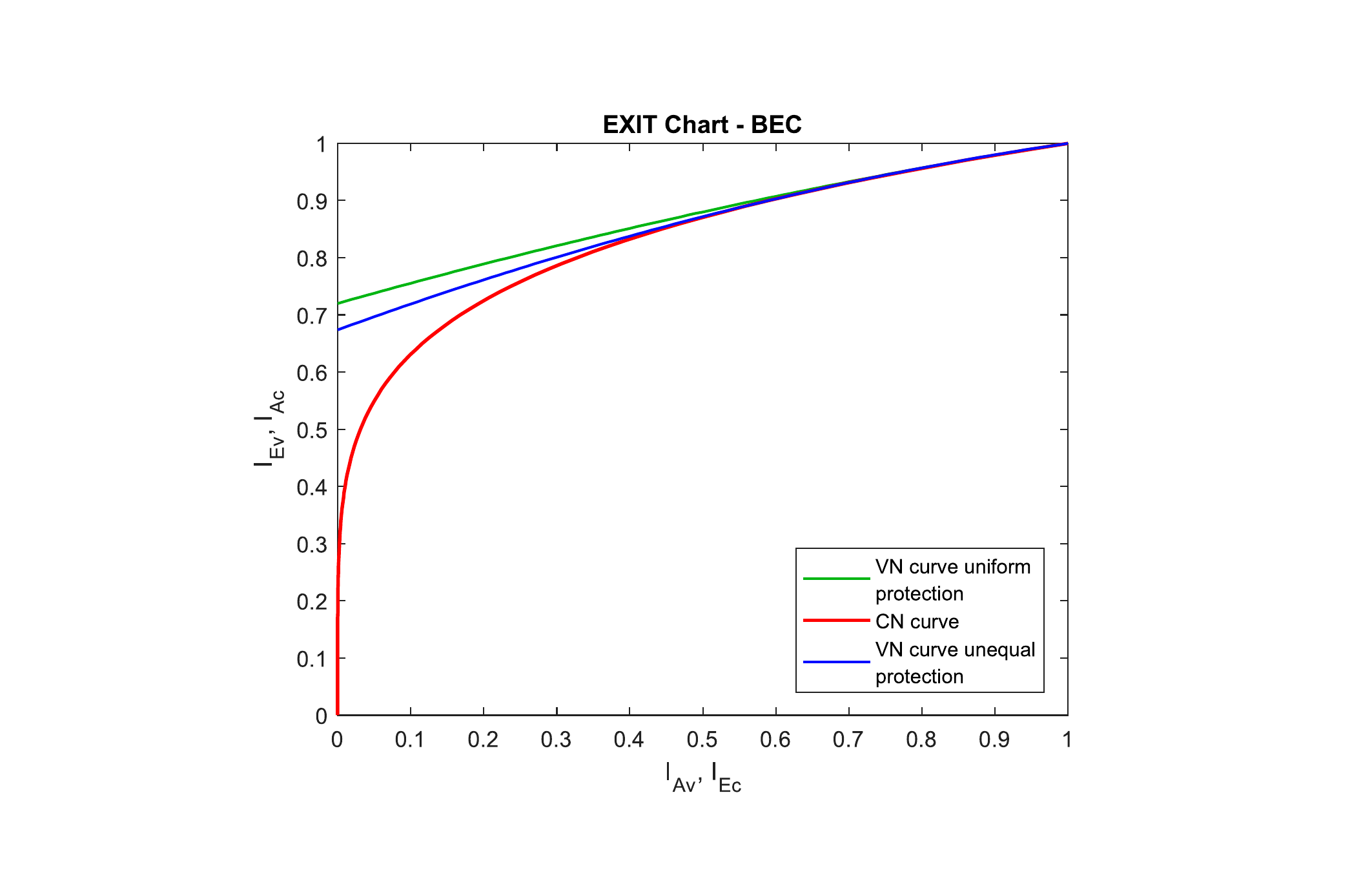}
\vspace{-1.8em}
\caption{EXIT chart on BEC. The LP problem is solved for uniform protection with $q_{\textup{uniform}}=0.28$, and the top curve represents $I_{\textup{E,v}}(I_{\textup{A}})$. For the same code parameters, $q'_1 = 0.003$ and $q'_2 = 0.488,$ are reached for UEP, and the middle curve represents $I^{*}_{\textup{E,v}}(I_{\textup{A}})$. The bottom curve represents $I^{-1}_{\textup{E,c}}(I_{\textup{A}})$.}
\label{fig_2}
\vspace{-1.5em}
\end{figure}

In Fig.~\ref{fig_2}, the top curve, which is the VN curve for uniform protection $I_{\textup{E,v}}(I_{\textup{A}})$, is generated by substituting the solution of the LP problem in (\ref{UEP_bec_fcn}) with $q_{\textup{uniform}} = 0.28$. The CN inverse curve $I^{-1}_{\textup{E,c}}(I_{\textup{A}})$, which is the bottom curve, is generated from (\ref{fcn_bec_Iec_inv}). As observed, LP constraints are satisfied. The middle curve, which is the VN curve with UEP $I^{*}_{\textup{E,v}}(I_{\textup{A}})$, is plotted by substituting the solution of the LP problem in (\ref{UEP_bec_fcn}) with the best $q'_1$ and $q'_2$ values that satisfy the constraints. Fig.~\ref{fig_2} also demonstrates the decoding convergence for the UEP setup (red/bottom curve is always below blue/middle curve).

In our second method of investigating the threshold gains, we do the opposite. We find the optimal degree distribution for given unequal erasure probabilities $q_1$ and $q_2$ with $q_{\textup{avg}}$ obtained as in (\ref{eqn_qavg}), where $q_1<q_2$. Next, we find the highest uniform erasure probability $q'_{\textup{uniform}} = q'_1 = q'_2$ satisfying the constraints given the code parameters. Numerical results demonstrating the UEP gains are illustrated in the right panel of Table~\ref{table1}.

Table~\ref{table1} presents the threshold gains we get by applying our UEP idea on BEC. For both methods, as the rate increases, i.e, as the system gets better (lower erasure probabilities), gains decrease as expected. The results of the first method, which optimizes for uniform protection, demonstrate that UEP offers a higher gain of around $10\%$ for a moderate rate of around $0.64$, which is suitable to wireless communication. There are also gains at higher rates suitable for data storage. This threshold gain means that with our UEP setup, the exact same code can appropriately operate at average erasure probabilities that are up to $10\%$ higher than the maximum it can operate at under the uniform setup. The second method optimizes for UEP instead of uniform protection, and it is included in this paper as a proof of concept since it slightly favors our UEP idea. The results of this method demonstrate a gain of around $17\%$ for a rate of around $0.64$. This is a further demonstration that UEP via more reliable parity bits is a promising idea that achieves significant threshold gains.

We note that BEC is the simplest channel to start with. The gains in Table~\ref{table1} motivate applying our UEP idea to error (bit flip) channels like BSC, which is discussed in the next section, and to more advanced data storage channels, which is left for our future work.

\section{Unequal Error Protection on BSC}\label{sec_bsc}

Let the communication and extrinsic channels in Fig.~\ref{fig_1} be BSCs with crossover probabilities $\epsilon$ and $\delta$, respectively. In this section, we apply the steps outlined in Section~\ref{sec_steps} and discuss the results of applying unequal error protection on parity and input information bits of the RA code for the BSC.

\noindent\textbf{\underline{Step~1:}} When considering the VNs of the RA code, $\underline{u} = \underline{x}$, $\underline{u}$ is one bit, and the Encoder is a repetition code with length $d_{\textup{v}}$. It is known that for a BSC with crossover probability $\delta$,
\begin{align}
    I_{\textup{A}} =  I(V_{1};A_{1}) = 1-H(\delta),   
\end{align}
where $H(\delta)$ is the binary entropy function with:
\begin{align}
    H(\delta) = -\delta\textup{log}_2(\delta) - (1-\delta)\textup{log}_2(1-\delta). \nonumber
\end{align}

We also use $H(X)$ to represent the entropy of a random variable $X$. Next, we derive $I_{\textup{E,v}}$ for VNs in Theorem~\ref{thm_bsc_Iev} for fixed $d_{\textup{v}}$.

\begin{theorem}\label{thm_bsc_Iev}
Consider the VNs of the RA code. The average extrinsic information coming out of decoder when the communication and extrinsic channels are BSCs with crossover probabilities $\epsilon$ and $\delta$, respectively, is:
\begin{align} \label{bsc_Iev}
I_{\textup{E,v}} &= 1 - \sum_{i=0}^{d_{\textup{v}}-1}\binom{d_{\textup{v}}-1}{i}\big[\theta^{i}_1+\theta^{i}_2\big]H\bigg(\frac{\theta^{i}_2}{\theta^{i}_1 + \theta^{i}_2}\bigg),
\end{align}
where
\begin{align} \label{bsc_Iev_terms}
    \theta^{i}_1 = (1-\epsilon)(1-\delta)^{i}\delta^{d_{\textup{v}}-1-i} \text{ and } \theta^{i}_2 = \epsilon \delta^{i}(1-\delta)^{d_{\textup{v}}-1-i}.
\end{align}

\end{theorem}

\begin{IEEEproof}
We prove Theorem~\ref{thm_bsc_Iev} by directly using (\ref{IE}) and calculating the mutual information:
\begin{align} \label{rule}
    I_{\textup{E,v}} &= I(V_{1};\underline{Y},\underline{A}_{[1]}) = H(V_{1})-H(V_{1} | \underline{Y},\underline{A}_{[1]}),
\end{align}
where $H(V_{1}) = 1$. Next, we calculate the conditional entropy:
\begin{align} \label{cond_entropy}
    H(V_{1} | \underline{Y},\underline{A}_{[1]}) &= - \sum_{v_1} \sum_{\underline{y}} \sum_{\underline{a}_{[1]}} p(v_{1})p(\underline{y},\underline{a}_{[1]} | v_{1}) \nonumber \\ &\hspace{+1.0em} \cdot \textup{log}_2\bigg(\frac{p(v_{1})p(\underline{y},\underline{a}_{[1]} | v_{1})}{p(\underline{y},\underline{a}_{[1]})}\bigg),
\end{align}
and $p(\underline{y},\underline{a}_{[1]} | v_{1}) = p(\underline{y}| v_{1})p(\underline{a}_{[1]} | v_{1})$.

We start with:
\begin{align} \label{h1}
    p(v_{1}=0) = p(v_{1}=1) = \frac{1}{2}.
\end{align}
Additionally, we know that:
\begin{align} 
    &p(\underline{y}= 0| v_{1}=0) = p(\underline{y}= 1| v_{1}=1) = (1-\epsilon), \label{h2}\\
    &p(\underline{y}= 0| v_{1}=1) = p(\underline{y}= 1| v_{1}=0) = \epsilon. \label{h22}
\end{align}

Next, recall that $\underline{a}_{[1]}$ is a binary vector of length $d_{\textup{v}}-1$. Let the superscript $i$ in $\underline{a}^{i}_{[1]}$ indicate that among the $d_{\textup{v}}-1$ entries of $\underline{a}^{i}_{[1]}$, $i$ of them are $0$'s and $d_{\textup{v}}-1-i$ of them are $1$'s. For example, $\underline{1} = \underline{a}^{0}_{[1]}$ and $\underline{0} = \underline{a}^{d_{\textup{v}}-1}_{[1]}$. Observe that due to BSC, $p(\underline{a}^{i}_{[1]}| v_{1}=0) = p(\underline{a}^{d_{\textup{v}}-1-i}_{[1]}| v_{1}=1)$. Thus, for general $i \in \{0,1,2,\dots,d_{\textup{v}}-1 \}$,
\begin{align} \label{h3}
    p(\underline{a}^{i}_{[1]} | v_{1}=0) = (1-\delta)^{i}\delta^{d_{\textup{v}}-1-i} = p(\underline{a}^{d_{\textup{v}}-1-i}_{[1]} | v_{1}=1) .
\end{align}

Next, we calculate the joint probability term $p(\underline{y},\underline{a}_{[1]})$ in (\ref{cond_entropy}). We have:
\begin{equation} \label{ex2}
    p(\underline{y},\underline{a}^{i}_{[1]}) = \sum_{v_1}  p(\underline{y},\underline{a}^{i}_{[1]} | v_1) p(v_1).
\end{equation}
Observe that due to BSC , $p(\underline{y}= 0,\underline{a}^{i}_{[1]}) = p(\underline{y}= 1,\underline{a}^{d_{\textup{v}}-1-i}_{[1]})$. Thus, for general $i \in \{0,1,2,\dots,d_{\textup{v}}-1 \}$,
\begin{align} \label{h4}
    p(\underline{y}= 0,\underline{a}^{i}_{[1]}) &= \frac{1}{2}\big[(1-\epsilon)(1-\delta)^{i}\delta^{d_{\textup{v}}-1-i}  \nonumber \\ &\hspace{+1.0em} + \epsilon\delta^{i}(1-\delta)^{d_{\textup{v}}-1-i}\big] = p(\underline{y}= 1,\underline{a}^{d_{\textup{v}}-1-i}_{[1]}),
\end{align}
which provides the joint probability for all pairs of $\underline{y}$ and $\underline{a}_{[1]}$. 

Now, we substitute equations (\ref{h1}), (\ref{h2}), (\ref{h22}), (\ref{h3}), and (\ref{h4}) in (\ref{cond_entropy}). Note that for each $i$, there exist $\binom{d_{\textup{v}}-1}{i}$ distinct $\underline{a}^{i}_{[1]}$ vectors. For example, if we substitute $(v_{1}=0, \underline{y}=0, \underline{a}^{0}_{[1]})$ and $(v_{1}=0, \underline{y}=1, \underline{a}^{d_{\textup{v}}-1}_{[1]})$ in the summations of (\ref{cond_entropy}), we get:
\begin{align} \label{ex3}
    \frac{1}{2}\binom{d_{\textup{v}}-1}{0}\big[\theta^{0}_1+\theta^{0}_2\big]H\bigg(\frac{\theta^{0}_2}{\theta^{0}_1 + \theta^{0}_2}\bigg) \nonumber
\end{align}
for the two scenarios combined, where
\begin{equation}
\theta^{0}_1 = (1-\epsilon)\delta^{d_{\textup{v}}-1} \text{ and } \theta^{0}_2 = \epsilon(1-\delta)^{d_{\textup{v}}-1} \nonumber.   
\end{equation}
When $v_{1}=1$, the above result is multiplied by $2$ due to symmetry. Finally, if we compute the summations for all values of $v_1$, $\underline{y}$, and $\underline{a}_{[1]}$, we obtain:
\begin{align} \label{cond}
    H(V_{1} | \underline{Y},\underline{A}_{[1]}) &= \sum_{i=0}^{d_{\textup{v}}-1}\binom{d_{\textup{v}}-1}{i}\big[\theta^{i}_1+\theta^{i}_2\big]H\bigg(\frac{\theta^{i}_2}{\theta^{i}_1 + \theta^{i}_2}\bigg),
\end{align}
where
\begin{align} \label{condd}
    \theta^{i}_1 = (1-\epsilon)(1-\delta)^{i}\delta^{d_{\textup{v}}-1-i} \text{ and } \theta^{i}_2 = \epsilon \delta^{i}(1-\delta)^{d_{\textup{v}}-1-i}.
\end{align}
Substituting (\ref{cond}) and (\ref{condd}) in (\ref{rule}), completes the proof.
\end{IEEEproof}

To derive the EXIT function $I_{\textup{E,v}}(I_{\textup{A}})$ in terms of $I_{\textup{A}}$, we use (\ref{bsc_Iev}) and (\ref{bsc_Iev_terms}) with $\delta = H^{-1}(1-I_{\textup{A}})$ for $\delta \in (0,0.5)$ ($ I_{\textup{A}} \in (0,1)$), where $H^{-1}$ is inverse binary entropy function.

\noindent \textbf{\underline{Step~2:}} When considering the CNs of the RA code, the switch on the top branch is open, and the Encoder is an SPC code with length $d_{\textup{c}}$. It is clear that:
\begin{align}
    I_{\textup{A}} =  I(V_{1};A_{1}) = 1-H(\delta).
\end{align}

We derive $I_{\textup{E,c}}$ for CNs in Theorem~\ref{thm_bsc_Iec} for fixed $d_{\textup{c}}$.
\begin{theorem}\label{thm_bsc_Iec}
Consider the CNs of the RA code. The average extrinsic information coming out of decoder when the communication and extrinsic channels are BSCs with crossover probabilities $\epsilon$ and $\delta$, respectively, is:
\begin{align} \label{bsc_Iec}
I_{\textup{E,c}} &= 1 - H\bigg(\frac{1}{2}\Big[1-(1-2\delta)^{d_{\textup{c}}-1}\Big]\bigg).
\end{align}

\end{theorem}

\begin{IEEEproof}
We prove Theorem~\ref{thm_bsc_Iec} by directly using (\ref{IE}) and calculating the mutual information. Recall that the switch on the top branch is open. Thus,
\begin{align} \label{rule2}
    I_{\textup{E,c}} &= I(V_{1};\underline{A}_{[1]}) = 1 -H(V_{1} | \underline{A}_{[1]}).
\end{align}
Next, we calculate the conditional entropy,
\vspace{-0.1em}\begin{align} \label{cond_entropy2}
    H(V_{1} | \underline{A}_{[1]}) \negthickspace &= - \negthickspace \sum_{v_1} \sum_{\underline{a}_{[1]}} p(v_{1})p(\underline{a}_{[1]} | v_{1})\textup{log}_2\Big(\frac{p(v_{1})p(\underline{a}_{[1]} | v_{1})}{p(\underline{a}_{[1]})}\Big).
\end{align}

We start with:
\begin{align} \label{h1_2}
    p(v_{1}=0) = p(v_{1}=1) = \frac{1}{2}.
\end{align}
Additionally, for any binary vector $\underline{a}_{[1]}$,
\begin{align} \label{h2_2}
   p(\underline{a}_{[1]}) = p(\underline{v}_{[1]}) = \frac{1}{2^{d_{\textup{c}}-1}}.
\end{align}

Let $E$ and $O$ be the sets of binary vectors of length $d_{\textup{c}}-1$ that have even and odd number of $1$'s, respectively. We know that to satisfy the SPC equation, $v_{1}=0$ implies that $\underline{v}_{[1]}$ has an even number of $1$'s, and $v_{1}=1$ implies that $\underline{v}_{[1]}$ has an odd number of $1$'s.\footnote{The event that the SPC equation is satisfied, i.e, $\underline{v}$ has an even number of $1$'s, holds by default. Additional conditioning on this event does not make any impact on the result.} Thus, for $\underline{a}_{[1]} \in E$,
\begin{align} 
    p(\underline{a}_{[1]} | v_{1} = 0) &= \sum_{\underline{v}_{[1]}} p(\underline{a}_{[1]} | \underline{v}_{[1]}, v_{1} = 0) p(\underline{v}_{[1]} | v_{1} = 0) \nonumber \\ &= \frac{1}{2^{d_{\textup{c}}-2}} \sum_{\underline{v}_{[1]}} p(\underline{a}_{[1]} | \underline{v}_{[1]}, v_{1} = 0). \label{lolo_1}
\end{align}
The summation at the end of (\ref{lolo_1}) is just the probability of an even number of flips at the extrinsic channel. Consequently,   
\begin{align}
    &p(\underline{a}_{[1]} | v_{1} = 0) = \frac{1}{2^{d_{\textup{c}}-2}} \hspace{-0.7em} \sum_{i=0}^{\left \lfloor {(d_{\textup{c}}-1)/2}\right \rfloor} \negthickspace \binom{d_{\textup{c}}-1}{2i} \delta^{2i}(1-\delta)^{d_{\textup{c}}-1-2i}, \label{h3_2}
\end{align}
Using the same logic for $\underline{a}_{[1]} \in O$ gives:
\begin{align}
    &p(\underline{a}_{[1]} | v_{1} = 0) = \frac{1}{2^{d_{\textup{c}}-2}} \hspace{-0.7em} \sum_{i=0}^{\left \lfloor {(d_{\textup{c}}-2)/2}\right \rfloor} \negthickspace \binom{d_{\textup{c}}-1}{2i+1} \delta^{2i+1}(1-\delta)^{d_{\textup{c}}-2-2i}. \label{h4_2}
\end{align}
Let $\underline{a}^{\textup{e}}_{[1]}$ be in $E$ and $\underline{a}^{\textup{o}}_{[1]}$ be in $O$. Observe that due to symmetry, $p(\underline{a}^{\textup{e}}_{[1]} | v_{1} = 0) = p(\underline{a}^{\textup{o}}_{[1]} | v_{1} = 1)$ and $p(\underline{a}^{\textup{o}}_{[1]} | v_{1} = 0) = p(\underline{a}^{\textup{e}}_{[1]} | v_{1} = 1)$. 

Now, we substitute equations (\ref{h1_2}), (\ref{h2_2}), (\ref{h3_2}), and (\ref{h4_2}) in (\ref{cond_entropy2}), and we get:
\begin{align}
    H(V_{1} | \underline{A}_{[1]}) &= H \bigg(\sum_{i=0}^{\left \lfloor {(d_{\textup{c}}-2)/2}\right \rfloor} \negthickspace \binom{d_{\textup{c}}-1}{2i+1} \delta^{2i+1}(1-\delta)^{d_{\textup{c}}-2-2i} \bigg) \nonumber \\ 
    &= H\bigg(\frac{1}{2}\Big[1-(1-2\delta)^{d_{\textup{c}}-1}\Big]\bigg) \label{cond2},
\end{align}
where second equation is obtained from Gallager's thesis \cite{gallager}. Finally, substituting (\ref{cond2}) in (\ref{rule2}) completes the proof.
\end{IEEEproof}

Continuing with this step, the EXIT function $I_{\textup{E,c}}(I_{\textup{A}})$ and the inverse EXIT function $I^{-1}_{\textup{E,c}}(I_{\textup{A}})$ for $ \delta \in (0,0.5)$ are then:
\begin{align}
     &I_{\textup{E,c}}(I_{\textup{A}}) = 1- H\bigg(\frac{1}{2}\Big[1-\Big(1-2H^{-1}(1- I_{\textup{A}})\Big)^{d_{\textup{c}}-1}\Big]\bigg), \\
     &I^{-1}_{\textup{E,c}}(I_{\textup{A}}) = 1- H\bigg(\frac{1}{2}\Big[1-\Big(1-2H^{-1}(1- I_{\textup{A}})\Big)^{\frac{1}{d_{\textup{c}}-1}}\Big]\bigg). \label{fcn_bsc_Iec_inv}
\end{align}

\begin{remark}
When the communication channel is BSC, the extrinsic channel can also be assumed as additive white Gaussian noise (AWGN) channel. In a way conceptually connected to the threshold analysis in \cite{gallager}, we derive primary conclusions by assuming the extrinsic channel as BSC in this work.
\end{remark}

\noindent\textbf{\underline{Step~3:}} We now apply the UEP setup explained in Section~\ref{sec_steps} to derive $I^{*}_{\textup{E,v}}$. Parity and input information bits transmitted through the communication channel face crossover probabilities $\epsilon_{1}$ and $\epsilon_{2}$, respectively. Substituting (\ref{bsc_Iev}), (\ref{bsc_Iev_terms}) with $\delta = H^{-1}(1-I_{\textup{A}})$, $\delta \in (0,0.5)$, in (\ref{general_Iev}), we derive the EXIT function $I^{*}_{\textup{E,v}}(I_{\textup{A}})$.

\noindent\textbf{\underline{Step~4:}} Substituting (\ref{fcn_bsc_Iec_inv}) and $I^{*}_{\textup{E,v}}(I_{\textup{A}})$ found in previous step in (\ref{constraint2}) completes the LP problem in (\ref{LP}). We now solve this LP problem to find the optimal degree distribution. That is, we maximize the code rate in (\ref{rate}), subject to degree distribution constraint (\ref{constraint1}) and EXIT convergence constraint (\ref{constraint2}).
\newline

We find the solution of the LP problem numerically using a software program. In order to investigate the threshold gains of our UEP idea over uniform protection, we applied the first method in BSC. As explained in detail in Section~\ref{sec_bec}, we first solve the LP problem for uniform protection case, i.e., $\epsilon_1 = \epsilon_2 = \epsilon_{\textup{uniform}}$, and find the optimal degree distribution. A code with this degree distribution has \textit{threshold} $\epsilon_{\textup{uniform}}$. For the same code construction, we search for the best ($\epsilon'_1, \epsilon'_2$) combination with the highest average that satisfies the EXIT convergence constraint in (\ref{constraint2}) using (\ref{fcn_bsc_Iec_inv}) and $I^{*}_{\textup{E,v}}(I_{\textup{A}})$ found in Step 3. The average crossover probability ($\epsilon'_{\textup{avg}}$) and the percentage threshold gains are computed as in (\ref{eqn_qavg}) and (\ref{eqn_tg}) by replacing $q$'s with $\epsilon$'s. Numerical results demonstrating the UEP threshold gains are provided in Table~\ref{table2}.

\begin{figure}
\vspace{-1.1em}
\centering
\hspace{+0.5em}\includegraphics[trim={1.3in 0.5in 1.3in 0.5in},width=3.4in]{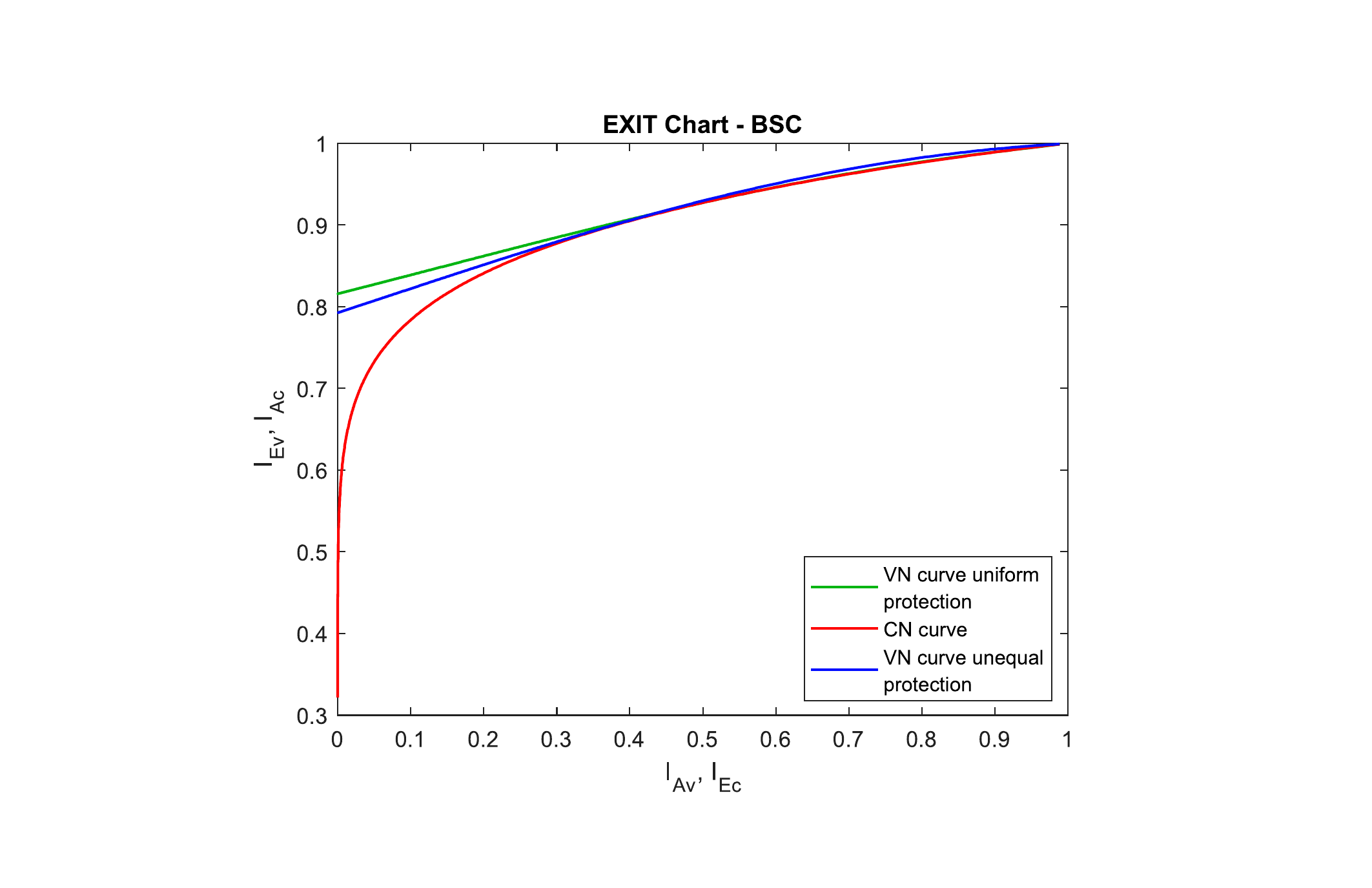}
\vspace{-1.9em}
\caption{EXIT chart on BSC. The LP problem is solved for uniform protection with $\epsilon_{\textup{uniform}}=0.028$, and the top curve represents $I_{\textup{E,v}}(I_{\textup{A}})$. For the same code parameters, $\epsilon'_1 = 0.001$ and $\epsilon'_2 = 0.039$ are reached for UEP, and the middle curve represents $I^{*}_{\textup{E,v}}(I_{\textup{A}})$. The bottom curve represents $I^{-1}_{\textup{E,c}}(I_{\textup{A}})$.}
\label{fig_3}
\vspace{-0.8em}
\end{figure}

In Fig.~\ref{fig_3}, the top curve, which is the VN curve for uniform error protection $I_{\textup{E,v}}(I_{\textup{A}})$, is generated by substituting the solution of the LP problem in $I^{*}_{\textup{E,v}}(I_{\textup{A}})$ with $\epsilon_{\textup{uniform}} = 0.028$. The CN inverse curve $I^{-1}_{\textup{E,c}}(I_{\textup{A}})$, which is the bottom curve, is generated from (\ref{fcn_bsc_Iec_inv}). As observed, LP constraints are satisfied. Using the solution of the LP problem, VN curve with UEP is generated, and it is the middle curve. As  observed, $I^{*}_{\textup{E,v}}(I_{\textup{A}})$ with the best $\epsilon'_1$ and $\epsilon'_2$ values ($\epsilon'_1 = 0.001, \epsilon'_2 = 0.039$), satisfies the constraints. 

\begin{table}
\caption{Threshold Gains of UEP at Various Crossover Probabilities When LP is Solved for Uniform Error Protection on BSC}
\vspace{-0.5em}
\centering

\begin{tabular}{|c|c|c|c|c|c|}
\hline
\multicolumn{6}{|c|}{\makecell{LP solved for uniform protection}} \\
\hline
Rate & $\epsilon_{\textup{uniform}}$ & $\epsilon'_1$ & {$\epsilon'_2$} & $\epsilon'_{\textup{avg}}$ & \textbf{Gain}   \\
\hline
$0.5815$ & $0.0820$ & $0.001$ & $0.1800$ & $0.1051$ & $28.2\%$ \\
\hline
$0.6349$ & $0.0670$ & $0.001$ & $0.1300$ & $0.0829$ & $23.7\%$ \\
\hline
$0.6747$ & $0.0570$ & $0.001$ & $0.1010$ & $0.0685$ & $20.1\%$ \\
\hline
$0.7051$ & $0.0498$ & $0.001$ & $0.0828$ & $0.0587$ & $17.8\%$ \\
\hline
$0.7305$ & $0.0440$ & $0.001$ & $0.0700$ & $0.0514$ & $16.8\%$ \\
\hline
$0.7525$ & $0.0390$ & $0.002$ & $0.0590$ & $0.0449$ & $15.1\%$  \\
\hline
$0.8032$ & $0.0280$ & $0.001$ & $0.0390$ & $0.0315$ & $12.6\%$  \\
\hline
$0.8545$ & $0.0180$ & $0.001$ & $0.0230$ & $0.0198$ & $10.0\%$ \\
\hline
\end{tabular}

\label{table2}
\vspace{-0.25em}
\end{table}

Table~\ref{table2} presents the threshold gains we obtain by applying our  UEP idea on BSC, when the LP problem is solved for uniform protection.\footnote{Note that for uniform protection, threshold lines are $3-13\%$ away from capacity lines (see also \cite{brink_ra} and \cite{verdu_ira}).} For the same code construction, the best unequal crossover probabilities satisfying the convergence constraints are found. In a way similar to Table~\ref{table1}, at moderate and lower rates corresponding to higher crossover probabilities, we obtain higher gains. At a rate of around $0.58$, which is suitable to wireless communication, $28\%$ gain is obtained with higher protection of parity bits. At data storage rates of around $0.75$ and $0.85$, significant gains of $15\%$ and $10\%$ are achieved, respectively. At a moderate rate of around $0.7$, almost $18\%$ gain is obtained. This gain means that with our UEP setup, the exact same code can appropriately operate at average crossover probabilities that are up to $18\%$ higher than the maximum it can operate under uniform setup. In other words, the threshold is improved by $18\%$.

In data storage terms, as the device deteriorates, lower rate codes are typically used to combat higher error probabilities and increase lifetime. Whereas with our UEP idea, lifetime gains can be achieved with no (or limited) rate loss.

\begin{remark}
In \cite{ahh_loco}, density gains of up to $20\%$ in MR channels were achieved when parity bits of an LDPC code are protected by constrained (LOCO) coding. We model the higher protection of parity bits via constrained coding by applying lower crossover (resp., erasure) probability to parity bits in BSC (resp., BEC). The threshold gains illustrated in Table~\ref{table2} (resp., Table~\ref{table1}) are consistent with, and thus confirm, the empirical results in \cite{ahh_loco}. This idea works because of the message passing decoding of LDPC codes. Highly reliable messages, e.g., LLRs, of parity bits are spread to other bits during message passing, which results in threshold gains in our UEP setup. In our future work, we will also simulate finite-length RA codes in BEC and BSC models to further investigate our UEP idea. We will also investigate this UEP idea for advanced channels, including the AWGN channel and channels with interference like MR \cite{ahh_loco}, Flash \cite{ahh_qaloco}, and two-dimensional MR (TDMR) channels \cite{ahh_general, ahh_md}.
\end{remark}

\section{Conclusion}\label{sec_conc}

We analyzed threshold gains of applying UEP via parity bits of an LDPC code on BEC and BSC. We modeled UEP using lower erasure/crossover probabilities for parity bits compared to input information bits on BEC/BSC for EXIT analysis. We used EXIT functions as a tool to understand the effect of a change in the mutual information of parity bits on the behavior of the overall coding scheme. We described the decoding model we use for EXIT analysis, and constructed a suitable RA code. We proposed a systematic methodology for the EXIT analysis of our UEP setup when both the communication and extrinsic channels are BEC/BSC. We derived a-priori and extrinsic information functions for VNs and CNs. For both channels, we constructed an LP problem to maximize the rate of the code under code construction and convergence constraints by finding the optimal degree distribution. After determining the code parameters, we made comparisons between the uniform protection setup and our UEP setup for the same code. We demonstrated up to around $17\%$ and $28\%$ UEP threshold gains in BEC and BSC, respectively. We plan to extend this work to AWGN channel along with practical data storage channels in our future work. We suggest that this UEP setup can contribute to significant density/lifetime gains in various data storage devices.

\section*{Acknowledgment}\label{sec_ack}

This research was supported by NSF under Grant CCF 1908730 and by AFOSR under Grant FA 8750-20-2-0504.


\end{document}